\documentclass[fullpaper]{jpsj2}

\def\ds{\displaystyle}

\def\i.e.{{\rm i.e.}}
\def\I{{\rm i}} 
\def\Ident{1}   
\def\System{{\rm S}}
\def\Bath{{\rm B}}
\def\SB{{\System\Bath}}
\def\IS{\Ident_\System}
\def\IB{\Ident_\Bath}
\def\e#1{{\rm e}^{#1}} 
\def\TrB#1{{\rm Tr}_\Bath\left\{#1\right\}}
\def\deltaFunction{{\rm \delta}}
\def\PShift{{\rm \Delta \theta}}
\def\Real{\,{\rm Re\!}}
\def\Imaginary{\,{\rm Im\!}}
\def\x{{\it x}}
\def\y{{\it y}}
\def\z{{\it z}}
\def\tv{\z}          
\def\lt{{\tiny \pm}} 
\def\H{{\cal H}} 
\def\hH{{\hat{\cal H}}} 
\def\HWt{\tilde\H}
\def\HS{\H_\System}
\def\HB{\H_\Bath}
\def\HSB{\H_\SB}
\def\tvH{\H_\tv}
\def\ltH{\H_\lt}
\def\Hp#1{\H_{\rm p}(#1)}
\def\HpH#1{\hH_{\rm p}(#1)} 
\def\Hbb#1{\H_{\rm bb}(#1)}
\def\Hbp#1{\H_{\rm bp}(#1)}

\def\HSBt{\HWt_\SB}
\def\HSBH#1{\hH_\SB(#1)}
\def\tvHSBH#1{\hH_\tv(#1)}
\def\ltHSBH#1{\hH_\lt(#1)}
\def\lSB#1{L_1 (#1)}

\def\LSBt#1{\tilde L_\SB(#1)}
\def\k{k}          
\def\kB{{k_\Bath}} 
\def\o{{\omega}}
\def\ok{{\o_\k}}
\def\gtk{g_{\k\tv}}    
\def\glk{g_{\k\lt}}    
\def\Cut{{\o_{\rm c}}}
\def\tG{G_{\!\tv}}
\def\lG{G_{\!\lt}}
\def\tTimeNon{{\tau_{\rm c}}}
\def\ttTimeNon{{\tau_{\!{\rm c} \tv}}}
\def\tlTimeNon{{\tau_{\!{\rm c} \lt}}}

\def\tTimeNonRatio{{\tlTimeNon/\ttTimeNon}}
\def\Dt{{\Delta t}}

\def\Nx#1{{N_{\!\x}(#1)}}
\def\Nz#1{{N_{\!\z}(#1)}}
\def\ihb{{\frac{\I}{\hbar}}}
\def\s{{\sigma}} 
\def\sx{{\s_\x}}
\def\sy{{\s_\y}}
\def\sz{{\s_\z}}
\def\sP{{\s_+}}
\def\sM{{\s_-}}

\def\sPMNx#1{{\s_{\rm b}(#1)}}
\def\sMPNx#1{{\s_{\rm b}^\dagger(#1)}}
\def\rW#1{\rho(#1)}
\def\rB{\rho_\Bath}
\def\rWt#1{{\tilde\rho}(#1)}
\def\rSt#1{{\tilde\rho}_\System(#1)}
\def\rBt#1{{\tilde\rho}_\Bath(#1)}
\def\rt#1{{\tilde\rho}_{#1}}

\def\2nB1#1{\coth\left(\frac{\hbar#1}{2\kB\!T}\right)}
\def\TO{{\it T}}       
\def\aT{\TO_{\rm\!a}}  
\def\TOe#1{\TO\e{#1}}  
\def\aTe#1{\aT\e{#1}}  
\def\anni{b_\k}        
\def\crea{b_\k^\dagger}
\def\roundy#1{\frac{\partial}{\partial#1}}
\def\Ket#1{\lvert#1\rangle}     
\def\abs#1{\lvert#1\lvert}      
\def\sumk{\sum_\k}
\def\intPM#1
{
 \int \hspace{-1mm}dt_1\hspace{-7mm}
  \begin{minipage}{8mm}\vspace{5.5mm}\tiny$(\!#1,\!t\!)$\end{minipage}
}

\def\Gauss#1{\left[#1\right]}
\def\commutator#1#2{[#1,#2]}
\def\tSpectral{I_{\!\tv}(\o)}
\def\lSpectral{I_{\!\lt}(\o)}

\title
{Pulse Control of Decoherence with Population Decay}

\author
{
Takahiro {\sc Murakami}\thanks{E-mail address: m1279028@hiroshima-u.ac.jp}
and Yositake {\sc Takane}
}

\inst
{
Department of Quantum Matter, ADSM, Hiroshima University,
1-3-1 Kagamiyama, Higashi-Hiroshima, Hiroshima 739-8530
}

\recdate{\hspace{15mm}}

\abst
{
The pulse control of decoherence in a qubit
interacting with a quantum environment
is studied with focus on a general case
where decoherence is induced by both pure dephasing and population decay.
To observe how the decoherence is suppressed by periodic $\pi$ pulses,
we present a simple method to calculate the time evolution of a qubit
under arbitrary pulse sequences consisting of bit-flips and/or phase-flips.
We examine the effectiveness of the two typical sequences:
bb sequence consisting of only bit-flips,
and bp sequence consisting of both bit- and phase-flips.
It is shown that the effectiveness of the pulse sequences
depends on a relative strength of the two decoherence processes
especially when a pulse interval is slightly shorter than
qubit-environment correlation times.
In the short-interval limit, however,
the bp sequence is always more effective than, or at least as effective as,
the bb sequence.
}

\kword{decoherence, pure dephasing, population decay, qubit}

\begin{document}

\maketitle
\section{Introduction}
To employ a two-level quantum system as a qubit for quantum computations,
a quantum superposed state prepared in it must be well preserved
for a long time.
However, since any quantum systems interact with their surrounding environment,
a superposed state is degraded by getting entangled with an environment.
The degradation of a quantum superposed state is called decoherence.
To perform reliable quantum computations,
we thus need to suppress decoherence caused by an environment.
Several schemes to suppress decoherence have been proposed so far
\cite
 {TombesiVitali,Shor,Steane,DuanGuo1,ZanardiRasetti,Ban,ViolaLloyd,agarwal}.
Among them, we focus on the decoherence suppression scheme
which uses a sequence of short $\pi$ pulses
\cite
{
 Ban,ViolaLloyd,DuanGuo,ViolaKnillLloyd,VitaliTombesi,UchiyamaAihara,
 PPHS,MurakamiTakane,TakaneMurakami
}
since it is relatively easy to implement in experimental situations
and can be readily applied to multi-qubit systems.
We call this scheme the pulse control of decoherence.
The pulse control has been proposed by Ban \cite{Ban},
and by Viola and Lloyd \cite{ViolaLloyd}.
They considered a qubit coupled with an environment
assuming that qubit-environment interactions induce decoherence
without population decay (\i.e., pure dephasing).
They calculated the time evolution of a qubit
under periodic $\pi$ pulses,
and showed that the pure dephasing can be suppressed
if the pulse interval is much shorter than the correlation time
for qubit-environment interactions.
It has been proposed that we can even suppress the decoherence
with population decay by applying $\pi$ pulses
\cite{DuanGuo,ViolaKnillLloyd}.

Let us consider a qubit linearly interacting with a boson environment.
The Hamiltonian is $\H_0 = \HS +\HB +\HSB$ with
\begin{eqnarray}
\HS &= &\hbar\o_0\frac{\sz}{2} \otimes \IB
\label{HS},
\\
\HB &= &\IS \otimes \sumk \hbar\ok \crea \anni
\label{HB}.
\end{eqnarray}
Here, $\HS$ and $\HB$ describe the qubit and the environment, respectively,
and $\HSB$ describes their mutual interactions.
We have used the pseudo-spin representation in expressing the qubit.
Ban, and Viola and Lloyd, considered the case of $\HSB=\tvH$
with
\begin{equation}
\tvH =\sumk
        \hbar \left(
               \gtk^* \sz \otimes \crea
               +\gtk  \sz \otimes \anni
              \right)
\label{HZ},
\end{equation}
which induces the pure dephasing
\cite{Ban,ViolaLloyd}.
They pointed out that the dephasing can be suppressed
by the sequence of periodic $\pi$ pulses
about the $\x$-axis in the pseudo-spin space.
They confirmed that the dephasing is notably suppressed
if the interval $\Dt$ between $\pi$ pulses is much shorter than
the correlation time for qubit-environment interactions.
Uchiyama and Aihara \cite{UchiyamaAihara} extended the argument
to a nonlinear interaction case by using a diagrammatic perturbation method.
However, we should note that the decoherence arises
from not only pure dephasing but also population decay.
A simple but standard model
which can describe the decoherence with population decay
is given by $\HSB=\ltH$ with
\begin{equation}
\ltH =\sumk
        \hbar \left(
               \glk^* \sM \otimes \crea
               +\glk  \sP \otimes \anni
              \right)
\label{Hpm}.
\end{equation}
It has been proposed that decoherence as well as population decay
can be suppressed by applying periodic $\pi$ pulses about the $\z$-axis
if the interval $\Dt$ is much shorter than the correlation time
\cite{DuanGuo,ViolaKnillLloyd}.
The present authors \cite{MurakamiTakane} confirmed this
by numerically calculating the time evolution
of the reduced density matrix for a qubit
based on the time-convolutionless projection operator approach.
Protopopescu \cite{PPHS} reported a numerical study
on the subject similar to this.
However, they employed the approximation
in which the commutator $\commutator{\Hp{t}}{\ltH}$
($\Hp{t}$: external pulse field) is completely neglected.
We have no reason to justify the approximation.

We focus on the case with
\begin{equation}
\HSB =\tvH+\ltH
\label{HSB}.
\end{equation}
This case has not been studied extensively.
This interaction Hamiltonian $\HSB$ induces
not only the pure dephasing but also the decoherence with population decay,
and describes more general situations than those treated
in refs. \citenum{Ban}, \citenum{ViolaLloyd} and \citenum{MurakamiTakane}.
Each $\pi$ pulse about the $\x$-axis ($\z$-axis)
induces a bit-flip (phase-flip).
Periodic bit-flips cancel out the influence of $\tvH$,
while periodic phase-flips cancel out the influence of $\ltH$.
Let $\ttTimeNon$ and $\tlTimeNon$ be the correlation times
for $\tvH$ and $\ltH$, respectively.
It has been shown that
if $\Dt$ is much shorter than $\ttTimeNon$ and $\tlTimeNon$,
the decoherence caused by $\HSB=\tvH+\ltH$ can be suppressed
by a sequence of both bit-flips and phase-flips
\cite{DuanGuo,ViolaKnillLloyd}.
The simplest sequence is composed
by the alternate application of a bit-flip and a phase-flip.
We call it bp sequence.
Although its effectiveness has been proved
in the short interval limit of $\Dt \ll \ttTimeNon$ and $\tlTimeNon$,
a realization of this condition is not necessarily possible
in actual experiments.
It is of interest to study the effectiveness
in experimentally relevant cases of $\Dt\lesssim\ttTimeNon$ and $\tlTimeNon$. 
Furthermore, we are also interested
in whether the bp sequence is always more effective
than the sequence consisting of only bit-flips.
We call the latter one bb sequence hereafter.
If $\ttTimeNon \sim \tlTimeNon$,
we expect that the bp sequence is much more effective than the bb sequence.
However, in the regime of $\ttTimeNon \ll \tlTimeNon$,
the bb sequence may predominate the bp sequence.

In this paper, we study the time evolution of a qubit
coupled with a quantum environment under periodic $\pi$ pulses,
to examine the effectiveness of the bp and bb sequences
when the interaction Hamiltonian is given by eq. (\ref{HSB}).
It is assumed that qubit-environment coupling is week.
We employ the time-convolutionless projection operator approach
following ref. \citenum{MurakamiTakane},
and derive the equation of motion
which describes the time evolution of the reduced density matrix for a qubit.
Our approach is simpler than that presented in ref. \citenum{UchiyamaAihara},
and more accurate than that presented in ref. \citenum{PPHS}.
Besides the approaches in refs. \citenum{UchiyamaAihara} and \citenum{PPHS},
several approximate methods based on the influence-functional formalism
may be applicable \cite{Weiss}.
However, such methods are too complicated for our restricted purpose.
The resulting equation has an advantage
that it is applicable to arbitrary pulse sequences
consisting of bit- and/or phase flips.
We examine the three cases of $\tTimeNonRatio=2$, $5$ and $50$
for a fixed value of $\tTimeNon$,
where $\tTimeNon$ is defined
by $\tTimeNon^{-2}\equiv\ttTimeNon^{-2} +\tlTimeNon^{-2}$.
The pulse interval is chosen as $\Dt/\tTimeNon=2^{-2}$, $2^{-3}$ and $2^{-4}$.
We show that the bp sequence is much more effective than the bb sequence
when $\tTimeNonRatio=2$.
However, we also show that their relative effectiveness can be reversed
in the cases of $\tTimeNonRatio=5$ and $50$
depending on $\Dt/\tTimeNon$.
Only in the short-interval limit, the bp sequence is
always more effective than, or at least as effective as, the bb sequence.
We conclude that the effectiveness of the pulse sequences depends on a
relative strength of the two decoherence processes when a pulse interval
is slightly shorter than $\tTimeNon$.


\section{Formulation}
We derive the equation of motion for the reduced density matrix for a qubit.
Let us consider the case
where the qubit-environment interaction is given by eq. (\ref{HSB})
and periodic $\pi$ pulses with an interval $\Dt$ are applied.
We also assume that the duration of each pulse is infinitely short.
The bb sequence consisting of only bit-flips
is described by
\begin{equation}
\Hbb{t}
=\hbar\frac{\pi}{2}
  \sum_{j=1}^\infty
   \deltaFunction(t-j\Dt)(\sx\cos\o_0t+\sy\sin\o_0t)   \otimes \IB
\label{Hbb}.
\end{equation}
The bp sequence consisting of both bit- and phase-flips
is described by
\begin{equation}
\Hbp{t}
=\hbar\frac{\pi}{2}
  \sum_{j=1}^\infty
   \left[
    \deltaFunction(t-(2j-1)\Dt)(\sx\cos\o_0t+\sy\sin\o_0t)
    +\deltaFunction(t-2j\Dt)\sz
   \right] \otimes \IB
\label{Hbp},
\end{equation}
where a bit-flip is applied at $t=(2j-1)\Dt$
while a phase-flip is applied at $t=2j\Dt$.
The total Hamiltonian is $\H(t) \equiv \H_0 +\Hp{t}$,
where $\Hp{t}$ represents either $\Hbb{t}$ or $\Hbp{t}$.
Although we treat only the bb and bp sequences,
the following argument can be applied to any pulse sequences
consisting of bit- and/or phase-flips.
Let $\rW{t}$ be the density matrix of the whole system.
It obeys the Liouville-von Neumann equation
\begin{equation}
\roundy{t}\rW{t} = L(t)\rW{t}
\label{LiouvilleEq.}
\end{equation}
with 
\begin{equation}
L(t) \,\cdot\, =-\ihb\commutator{\H(t)}{\,\cdot\,}
\label{LiouvilleOperator}.
\end{equation}
We decompose $\H(t)$ as $\H(t)=\H_1(t)+\HSB$,
where $\H_1(t)=\HS+\HB+\Hp{t}$.
It is convenient to introduce $\rWt{t}$ which is defined as
\begin{equation}
\rWt{t}=\aTe{-\int_0^t\!d\tau \lSB{\tau}} \rW{t},
\end{equation}
where $\aT$ is the anti-time-ordering operator
and $\lSB t \cdot = -\ihb\commutator{\H_1(t)}{\cdot}$.
From eq. (\ref{LiouvilleEq.}), we obtain
\begin{equation}
\roundy{t}\rWt{t} = \LSBt{t}\rWt{t}
\label{LiouvilleEq.t},
\end{equation}
where $\LSBt t \cdot = -\ihb\commutator{\HSBt(t)}{\cdot}$.
Here, $\HSBt(t)$ is given by
\begin{eqnarray}
\HSBt(t)
&=
&\aTe{\ihb\int_0^t\!d\tau \H_1(\tau)}
  \HSB \TOe{-\ihb\int_0^t\!d\tau \H_1(\tau)}
\nonumber
\\
&=
&\aTe{\ihb\int_0^t\!d\tau \HpH\tau}
  \HSBH{t}\TOe{-\ihb\int_0^t\!d\tau \HpH{\tau}}
\label{HSBt2},
\end{eqnarray}
where $\TO$ is the time-ordering operator,
$\HpH{t}=\e{\ihb(\HS+\HB)t}\Hp{t}\e{-\ihb(\HS+\HB)t}$ and
\begin{eqnarray}
\HSBH{t}
 &= &\e{\ihb(\HS+\HB)t} \HSB \e{-\ihb(\HS+\HB)t}
\nonumber
\\
&=
&\hbar
  \sumk
   \left(
     \gtk^* \sz \otimes \crea \e{ \I \ok      t}
     +\gtk  \sz \otimes \anni \e{-\I \ok      t}
   \right)
\nonumber \\ &&
 +\hbar
   \sumk
    \left(
     \glk^* \sM \otimes \crea \e{ \I(\ok-\o_0)t}
     +\glk  \sP \otimes \anni \e{-\I(\ok-\o_0)t}
    \right)
\label{HSBH}.
\end{eqnarray}
We can show that the time-ordered exponential factor in eq. (\ref{HSBt2}) is
obtained as
\begin{equation}
\TOe{-\ihb\int_0^t\!d\tau \HpH{\tau}}
=\left\{ \hspace{-1mm}
  \begin{array}{ll}\ds
   \IS \otimes \IB
     & \hspace{-2mm} \mbox{for }    0  < t <  \Dt ,
  \\ \ds
   \I\sx \otimes \IB
     & \hspace{-2mm} \mbox{for }  \Dt  < t < 2\Dt ,
  \\ \ds
   (\I\sx\times\I\sx) \otimes \IB
     & \hspace{-2mm} \mbox{for } 2\Dt  < t < 3\Dt ,
  \\ \ds
   (\I\sx\times\I\sx\times\I\sx) \otimes \IB
     & \hspace{-2mm} \mbox{for } 3\Dt  < t < 4\Dt ,
  \\ \ds
   \cdots
     & \cdots
  \end{array}
 \right.
\label{BBPulse}
\end{equation}
for $\Hp{t}=\Hbb{t}$ and
\begin{equation}
\TOe{-\ihb\int_0^t\!d\tau \HpH{\tau}}
=\left\{ \hspace{-1mm}
  \begin{array}{ll}\ds
   \IS \otimes \IB
     & \hspace{-2mm} \mbox{for }    0  < t <  \Dt ,
  \\ \ds
   \I\sx \otimes \IB
     & \hspace{-2mm} \mbox{for }  \Dt  < t < 2\Dt ,
  \\ \ds
   (\I\sz\times\I\sx) \otimes \IB
     & \hspace{-2mm} \mbox{for } 2\Dt  < t < 3\Dt ,
  \\ \ds
   (\I\sx\times\I\sz\times\I\sx) \otimes \IB
     & \hspace{-2mm} \mbox{for } 3\Dt  < t < 4\Dt ,
  \\ \ds
   \cdots
     & \cdots
  \end{array}
 \right.
\label{BPPulse}
\end{equation}
for $\Hp{t}=\Hbp{t}$.
Let $\Nx{t}$ ($\Nz{t}$) be the number of bit-flips (phase-flips)
within $[0,t]$.
If the bb sequence is applied (\i.e., $\Hp{t}=\Hbb{t}$), we find that
\begin{eqnarray}
\Nx{t} &= &\Gauss{\frac{t}{\Dt}},
\\
\Nz{t} &= &0 ,
\end{eqnarray}
where $\Gauss{a}$ denotes the integral part of the real number $a$.
In the case where the bp sequence is applied, we find that
\begin{eqnarray}
\Nx{t} &= &\Gauss{\frac{t+\Dt}{2\Dt}} ,
\\
\Nz{t} &= &\Gauss{\frac{t}{2\Dt}} .
\end{eqnarray}
If we define
\begin{equation}
\sPMNx{t}
=\left\{
  \begin{array}{ll}
   \sP & \mbox{if $\Nx{t}$ is even} ,
  \\
   \sM & \mbox{if $\Nx{t}$ is odd}  ,
  \end{array}
 \right.
\label{sPMNx}
\end{equation}
we obtain
\begin{equation}
\HSBt(t)
=(-1)^\Nx{t} \tvHSBH{t}   +(-1)^\Nz{t} \ltHSBH{t} ,
\label{nHSBt}
\end{equation}
where
\begin{equation}
\tvHSBH{t}
=\hbar
  \sumk
   \left(
    \gtk^*  \sz       \otimes \crea \e{ \I \ok      t}
    +\gtk   \sz       \otimes \anni \e{-\I \ok      t}
   \right)
\label{tvHSBH}
\end{equation}
and
\begin{equation}
\ltHSBH{t}
=\hbar
  \sumk
   \left(
    \glk^*  \sMPNx{t} \otimes \crea \e{ \I(\ok-\o_0)t}
    +\glk   \sPMNx{t} \otimes \anni \e{-\I(\ok-\o_0)t}
   \right)
\label{ltHSBH}.
\end{equation}
Equation (\ref{nHSBt}) holds for any pulse sequences
consisting of bit- and/or phase-flips,
although we have assumed that $\Hp{t}$ is equal to $\Hbb{t}$ or $\Hbp{t}$.
Note that each bit-flip $\pi$ pulse changes the sign of $\tvHSBH{t}$
and each phase-flip $\pi$ pulse changes the sign of $\ltHSBH{t}$.
The periodic sign change due to bit-flips (phase-flips)
effectively cancels out
the qubit-environment interaction $\tvHSBH{t}$ ($\ltHSBH{t}$).
This is the reason why the decoherence is suppressed
by the $\pi$ pulses.

Our interest is focused on the time evolution of the qubit,
which is described by the reduced density matrix
defined by
\begin{equation}
\rSt t =\TrB{ \rWt{t} }
=\begin{pmatrix}
  \rt{11}(t) & \rt{10}(t) \cr
  \rt{01}(t) & \rt{00}(t)
 \end{pmatrix} ,
\end{equation}
where
$\rt{11}$ ($\rt{00}$) represents the population in the upper (lower) state
and the off-diagonal terms characterize the coherence of the qubit.
We derive the equation of motion for $\rSt{t}$
following ref. \citenum{MurakamiTakane}.
In doing so, we assume that
the qubit and the environment are uncorrelated at $t=0$, \i.e.,
\begin{equation}
\rWt{0}=\rSt{0} \otimes \rBt{0}
\label{initial},
\end{equation}
and the environment is initially in thermal equilibrium
at temperature $T$, \i.e.,
\begin{equation}
\rBt{0}=\rB
       =\prod_\k \left( 1-\e{-\beta\hbar\ok} \right)
          \e{-\beta\hbar\ok \crea \anni }
\label{Bath}
\end{equation}
with $\beta=1/k_{\rm B}T$ ($k_{\rm B}$: the Boltzmann constant).
We introduce the projection operator $P$
which is defined as $P\cdot=\TrB{\cdot} \otimes\rB$.
We observe that $P\rWt{t}=\rSt{t}\otimes\rB$.
Adapting the time-convolutionless projection operator formalism
\cite{ShibataArimitsu}
with $P$ to eq. (\ref{LiouvilleEq.t}),
we derive the equation of motion for $\rSt{t}$,
\begin{equation}
\roundy{t}\rSt{t}
=\TrB{ \int_0^{t}dt_1 \LSBt{t}\LSBt{t_1} \bigl[\rSt{t}\otimes\rB\bigr] }
\label{SLiouvilleEq.t},
\end{equation}
where we have neglected higher-order terms by assuming that
the qubit-environment interaction is weak enough.
In calculating the right-hand side of eq. (\ref{SLiouvilleEq.t}),
we note that the decoherence without population decay
is mainly caused by the modes with $\ok\approx0$ in $\tvHSBH{t}$
and the decoherence with population decay is mainly caused by the modes
with $\ok\approx\o_0$ in $\ltHSBH{t}$.
Thus, we approximately neglect the cross terms
between $\tvHSBH{t}$ and $\ltHSBH{t}$ in eq. (\ref{SLiouvilleEq.t}).

Using eqs. (\ref{HSBH}) and (\ref{nHSBt}),
we obtain
\begin{eqnarray}
\roundy{t}\rt{11}(t) &= &-\gamma_{11}(t)\rt{11}(t)+\eta_{11}(t)
\label{DifferentialRho11} ,
\\
\roundy{t}\rt{10}(t)
&= &-\gamma_{10}^\Re(t) \Real\left[\rt{10}(t)\right]
    -\I\gamma_{10}^\Im(t) \Imaginary\left[\rt{10}(t)\right]
\label{DifferentialRho10} ,
\end{eqnarray}
where
\begin{eqnarray}
\gamma_{11}(t)
&= &2\sumk \abs{\glk}^2 \,\2nB1{\ok}
      \int_{0}^{t} dt_1\,(-1)^{\Nz{t}+\Nz{t_1}}
\nonumber \\ && \, \times
       \frac{1+(-1)^{\Nx{t}+\Nx{t_1}}}{2} \cos(\ok-\o_0)(t_1-t)
\label{gamma11} ,
\\
\eta_{11}(t)
&= &\sumk \abs{\glk}^2 \,
     \left[ \2nB1{\ok}-(-1)^\Nx{t} \right]
      \int_{0}^{t} dt_1\,(-1)^{\Nz{t}+\Nz{t_1}}
\nonumber \\ && \, \times
       \frac{1+(-1)^{\Nx{t}+\Nx{t_1}}}{2} \cos(\ok-\o_0)(t_1-t)
\label{eta11} ,
\\
\gamma_{10}^\Re(t)
&= &4\sumk \abs{\gtk}^2 \,\2nB1{\ok}
      \int_{0}^{t} dt_1\,(-1)^{\Nx{t}+\Nx{t_1}}
       \cos\ok(t_1-t)
\nonumber \\ && \,
    +\sumk \abs{\glk}^2 \,\2nB1{\ok}
      \int_{0}^{t} dt_1\,(-1)^{\Nz{t}+\Nz{t_1}}
\nonumber \\ && \, \times
       (-1)^{\Nx{t}+\Nx{t_1}} \e{\I(-1)^{\Nx{t}}(\ok-\o_0)(t_1-t)}
\label{gamma10Re} ,
\\
\gamma_{10}^\Im(t)
&= &4\sumk \abs{\gtk}^2 \,\2nB1{\ok}
      \int_{0}^{t} dt_1\,(-1)^{\Nx{t}+\Nx{t_1}}
       \cos\ok(t_1-t)
\nonumber \\ && \,
    +\sumk \abs{\glk}^2 \,\2nB1{\ok}
      \int_{0}^{t} dt_1\,(-1)^{\Nz{t}+\Nz{t_1}}
\nonumber \\ && \, \times
       \e{\I(-1)^{\Nx{t}}(\ok-\o_0)(t_1-t)}
\label{gamma10Im}.
\end{eqnarray}
In deriving eqs. (\ref{DifferentialRho11}) and (\ref{DifferentialRho10}),
we have used the two relations
$\rt{11}(t)+\rt{00}(t)=1$ and $\rt{10}(t)=\rt{01}^{\, *}(t)$.
Equations (\ref{DifferentialRho11}) and (\ref{DifferentialRho10})
describe the time evolution of the qubit under $\pi$ pulses.
It should be emphasized that the above equation can be applied
to any pulse sequences consisting of bit- and/or phase-flips.
Introducing the spectral functions $\tSpectral$ and $\lSpectral$ defined by
\begin{eqnarray}
\tSpectral &= &\sumk \abs{\gtk}^2 \deltaFunction(\o-\ok) ,
\\
\lSpectral &= &\sumk \abs{\glk}^2 \deltaFunction(\o-\ok) ,
\end{eqnarray}
we rewrite eq. (\ref{gamma10Im}) as
\begin{eqnarray}
\gamma_{10}^\Im(t)
&= &4\int_0^\infty d\o \tSpectral \2nB1{\o}
      \int_{0}^{t} dt_1\,(-1)^{\Nx{t}+\Nx{t_1}}
       \cos\o(t_1-t)
\nonumber \\ && \,
    +\int_0^\infty d\o \lSpectral \2nB1{\o}
      \int_{0}^{t} dt_1\,(-1)^{\Nz{t}+\Nz{t_1}}
       \e{\I(-1)^{\Nx{t}}(\o-\o_0)(t_1-t)}
\label{gamma10Spectral}.
\end{eqnarray}
Equations (\ref{gamma11}), (\ref{eta11}) and (\ref{gamma10Re})
are rewritten in the manner similar to eq. (\ref{gamma10Spectral}).

\section{Numerical Results}
\begin{figure}[tb]
{\Large(a)\hspace{-10mm}}
 \includegraphics*[scale=.63]{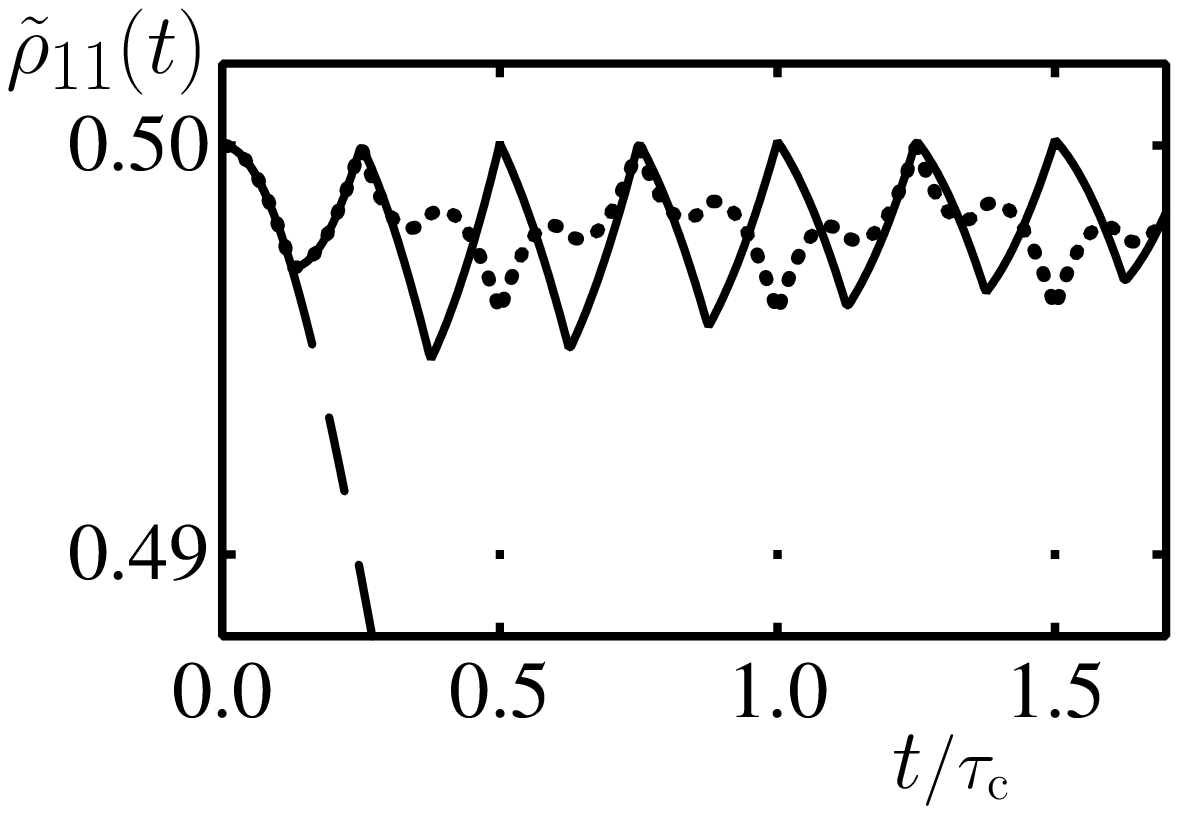}
\hspace{5mm}
{\Large(b)\hspace{-10mm}}
 \includegraphics*[scale=.63]{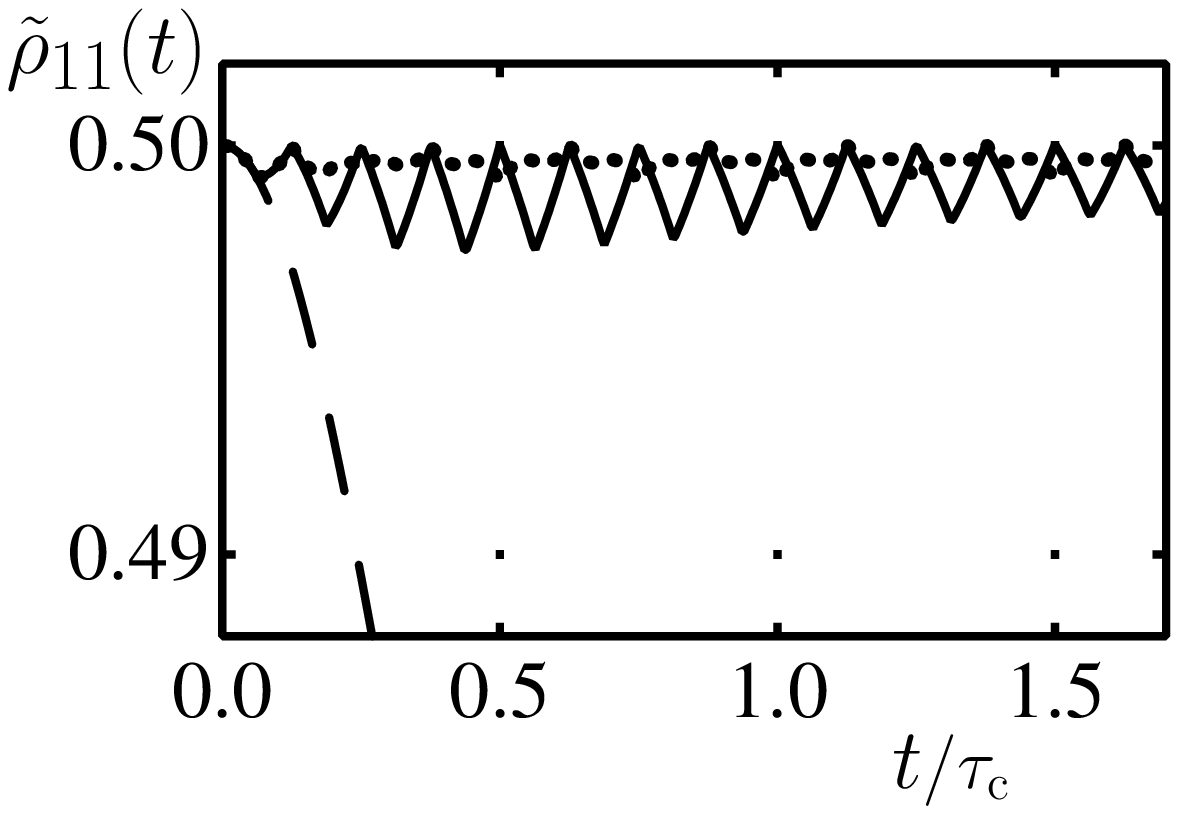}
\caption
 {
  Time evolution of $\rt{11}$ as a function of $t/\tTimeNon$
  at $\kB T / \hbar\Cut = 10^{-3}$.
  The ratio of the qubit-environment correlation times
  is $\tTimeNonRatio=2$.
  The pulse interval $\Dt$ is (a) $\Dt/\tTimeNon=2^{-3}$ and (b) $2^{-4}$.
  The solid lines and dotted lines correspond to the cases
  of the bb sequence and the bp sequence, respectively.
  The dashed lines represent the case where $\pi$ pulses are absent.
 }
\label{Population}
\end{figure}
On the basis of the resulting equation of motion,
we study the time evolution of the reduced density matrix $\rSt{t}$
in the presence of periodic $\pi$ pulses with an interval $\Dt$.
We numerically calculate $\rt{11}(t)$, $\abs{\rt{10}(t)}$ and $\PShift(t)$,
where $\PShift(t)$ is defined by
$\PShift(t)=\arg\{\rt{10}(t)\} -\arg\{\rt{10}(0)\}$.
We assume that the spectral function is given by
\begin{eqnarray}
\tSpectral &= &\tG \o \e{-\o/\Cut}
\label{tSpectral},
\\
\lSpectral &= &\lG \o \e{-\o/\Cut}
\label{lSpectral},
\end{eqnarray}
where $\Cut$ is a cut-off frequency,
and $\tG$ and $\lG$ are coupling constants.
The following parameters are employed:
$\o_0/\Cut=0.1$ and $\kB T/(\hbar\Cut)=0.001$.
In terms of the coupling constants,
the correlation times for $\tvH$ and $\ltH$ are
given by $\ttTimeNon \equiv (1/2\tG)^{1/2}\Cut^{-1}$
and $\tlTimeNon \equiv (2/\lG)^{1/2}\Cut^{-1}$, respectively.
The initial behavior of $\rt{10}(t)$ is expressed
as $\rt{10}(t)= \rt{10}(0)\exp[-(t/\tTimeNon)^2]$, where
\begin{equation}
\frac{1}{\tTimeNon^2}\equiv \frac{1}{\ttTimeNon^2}+\frac{1}{\tlTimeNon^2}.
\end{equation}
We fix $\tTimeNon$ at $\tTimeNon=0.4\times(2\pi/\Cut)$
and examine the cases of $\tTimeNonRatio=2$, $5$ and $50$.
While the pure dephasing is slightly stronger
than the decoherence with population decay
in the case of $\tTimeNonRatio=2$,
the former predominates the latter
in the case of $\tTimeNonRatio=50$.
For each case, we calculate the time evolution
with $\Dt/\tTimeNon=2^{-2}$, $2^{-3}$ and $2^{-4}$.
We choose $(\Ket{0}+\Ket{1})/\sqrt2$ and $(\Ket{0}+\I\Ket{1})/\sqrt2$
as initial states.
From numerical results, we find that
the decoherence suppression is relatively weak in the latter case.
Thus, we present only the results for $(\Ket{0}+\I\Ket{1})/\sqrt2$ with
\begin{equation}
\rSt0=\begin{pmatrix}
         1/2& \I/2\cr
       -\I/2&  1/2
      \end{pmatrix}.
\end{equation}

In Fig. \ref{Population}, we display $\rt{11}(t)$
with (a) $\Dt/\tTimeNon=2^{-3}$ and (b) $2^{-4}$
in the case of $\tTimeNonRatio=2$.
We do not display the results for $\tTimeNonRatio=5$ and $50$
because variations of $\rt{11}(t)$ are very small.
The decrease of $\rt{11}$ represents the population decay.
We observe that
the population decay is suppressed by either of the bb or bp sequences.
It should be emphasized again that
the bb sequence cannot suppress the influence of $\ltH$,
which induces population decay,
while the influence is suppressed by the bp sequence
containing phase-flip $\pi$ pulses.
The suppression of the population decay under the bb sequence
is simply attributed to periodic exchanges of the upper-state population
with the lower-state population due to bit-flip $\pi$ pulses.

\begin{figure}[t]
{\Large(a)\hspace{-10mm}}
 \includegraphics*[scale=.44]{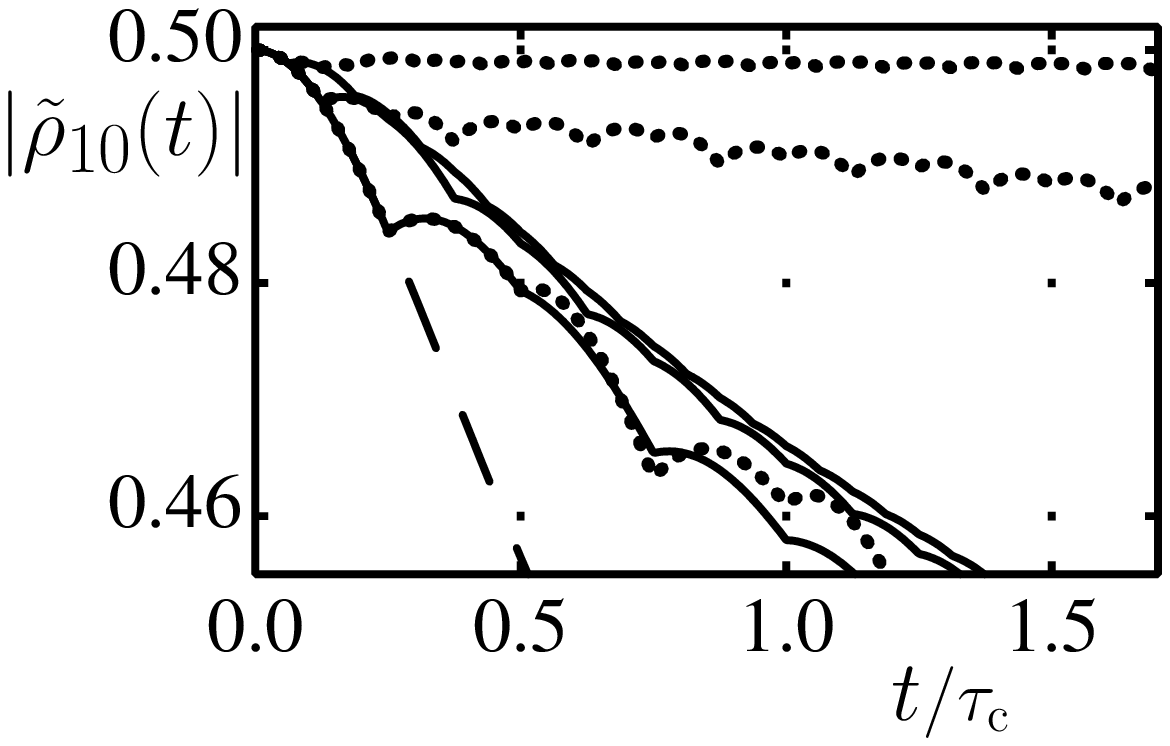}
{\Large(b)\hspace{-10mm}}
 \includegraphics*[scale=.44]{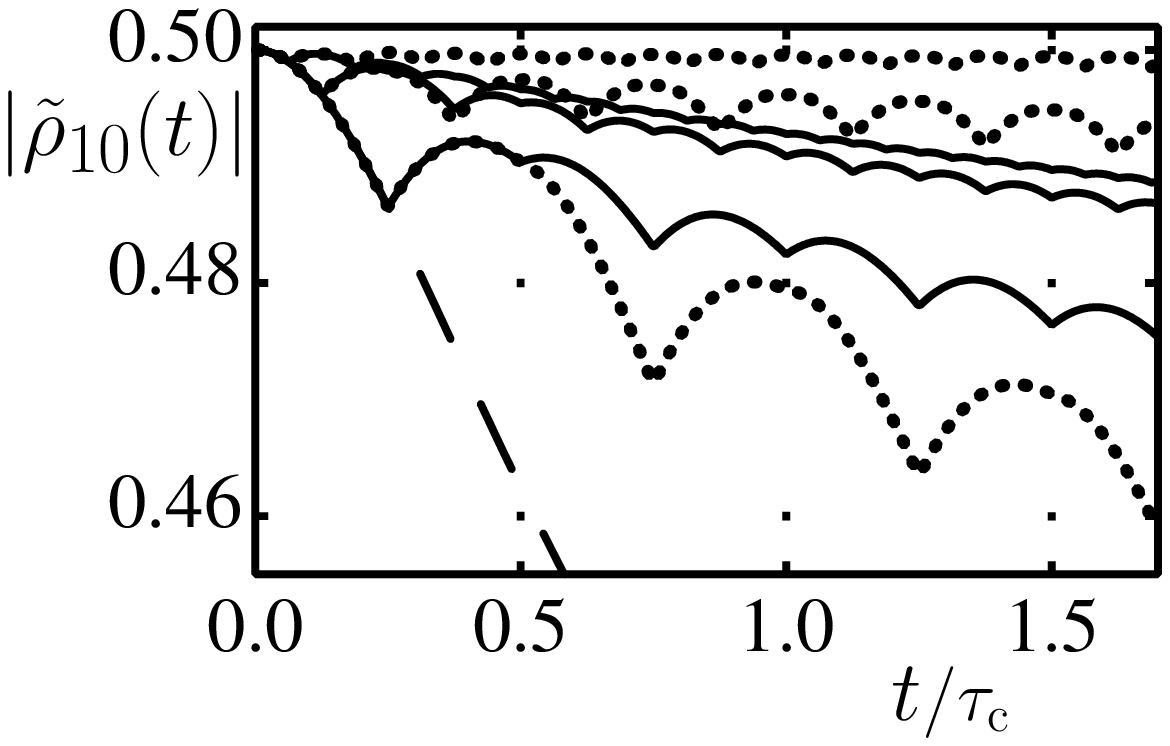}
{\hspace{1mm}\Large(c)\hspace{-10mm}}
 \includegraphics*[scale=.44]{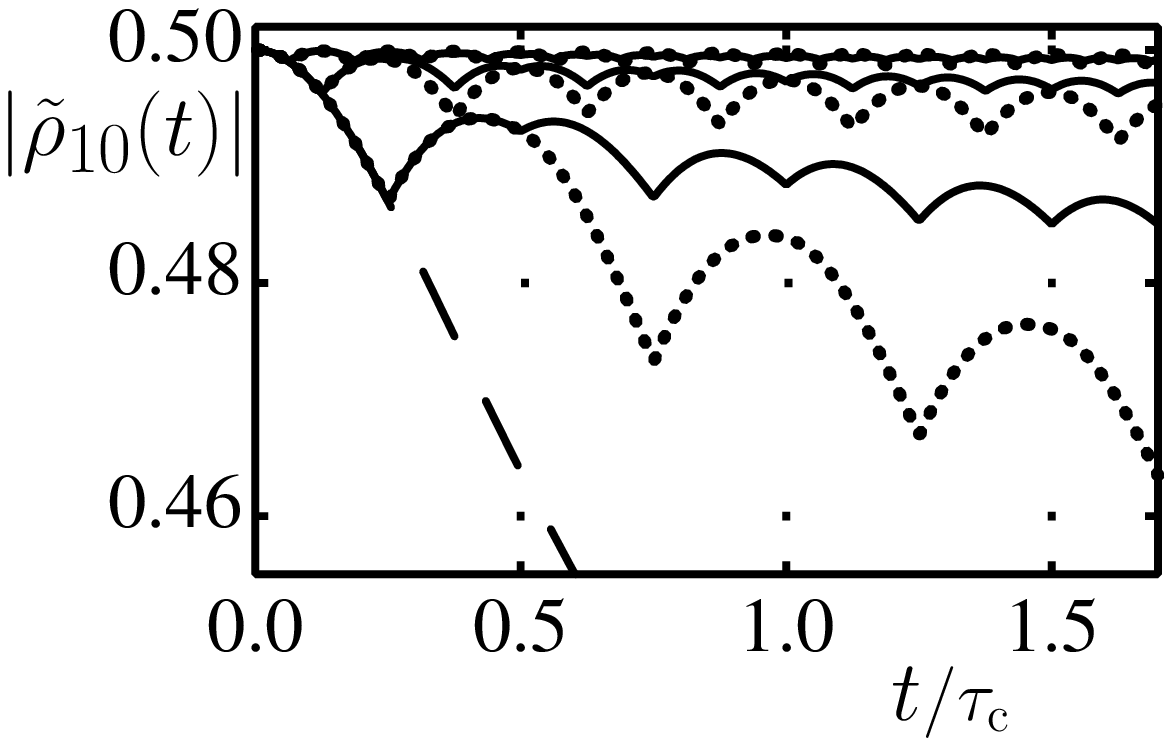}
\caption
 {
  Time evolution of $\abs{\rt{10}}$ 
  in the case of (a) $\tTimeNonRatio=2$, (b) $5$ and (c) $50$.
  The solid lines and dotted lines correspond to the cases
  of the bb sequence and the bp sequence, respectively,
  with $\Dt/\tTimeNon=2^{-2}$, $2^{-3}$ and $2^{-4}$
  from lower to upper lines.
  The dashed lines represent the case where $\pi$ pulses are absent.
 }
\label{Coherence}
\end{figure}
We compare $\rt{10}(t)$ under the bp sequence with that under the bb sequence.
In Fig. \ref{Coherence}, we display $\abs{\rt{10}(t)}$ with
$\Dt/\tTimeNon=2^{-2}$, $2^{-3}$ and $2^{-4}$
when (a) $\tTimeNonRatio=2$, (b) $5$ and (c) $50$.
The decrease of $\abs{\rt{10}(t)}$ represents the decoherence.
Figure \ref{Coherence} shows that
in the case of $\tTimeNonRatio=2$,
the decoherence suppression under the bp sequence
is much more effective than that under the bb sequence.
For the bb sequence,
the decrease of $\Dt$ from $\Dt/\tTimeNon=2^{-3}$ to $2^{-4}$
does not result in a notable improvement.
This fact is attributed to a shortcoming of the bb sequence
which cannot cancel out the influence of $\ltH$.
However, the effectiveness of the bb sequence is
improved with increasing $\tTimeNonRatio$
since the influence of $\ltH$ is relatively reduced.
In the case of $\tTimeNonRatio=5$,
the bb sequence is more effective
than the bp sequence for $\Dt/\tTimeNon=2^{-2}$,
and is predominated by the bp sequence
for $\Dt/\tTimeNon=2^{-3}$ and $2^{-4}$.
In the case of $\tTimeNonRatio=50$,
the bb sequence is more effective
than the bp sequence for $\Dt/\tTimeNon=2^{-2}$ and $2^{-3}$,
and is as effective as the bp sequence
in the shortest interval case of $\Dt/\tTimeNon=2^{-4}$.
The above results indicate that
the effectiveness of the pulse sequences
depends on a relative strength of the two decoherence processes
especially when a pulse interval is slightly shorter than $\tTimeNon$.
Only in the short-interval limit, the bp sequence is
more effective than, or at least as effective as,
the bb sequence regardless of $\tTimeNonRatio$.
%
%

The above results are explained as follows.
The bp sequence contains both phase-flips and bit-flips,
and thus can cancel out the influence of both $\tvH$ and $\ltH$,
while the bb sequence can cancel out the influence of only $\tvH$.
Thus, the bp sequence is much more effective than the bb sequence
in the case of $\tTimeNonRatio \sim 1$,
where the influence of both $\tvH$ and $\ltH$ is important.
However, the interval of two adjacent bit-flips in the bp sequence
is equal to $2\Dt$, and is twice longer than that in the bb sequence.
This indicates that the bp sequence reduces the influence of $\tvH$
less effectively than the bb sequence.
Thus, in the case of $\tTimeNonRatio\gg1$,
where $\tvH$ is more important than $\ltH$,
the bp sequence becomes less effective than the bb sequence
unless $\Dt$ is very much shorter than $\ttTimeNon$.

\begin{figure}[t]
{\Large(a)\hspace{-10mm}}
 \includegraphics*[scale=.63]{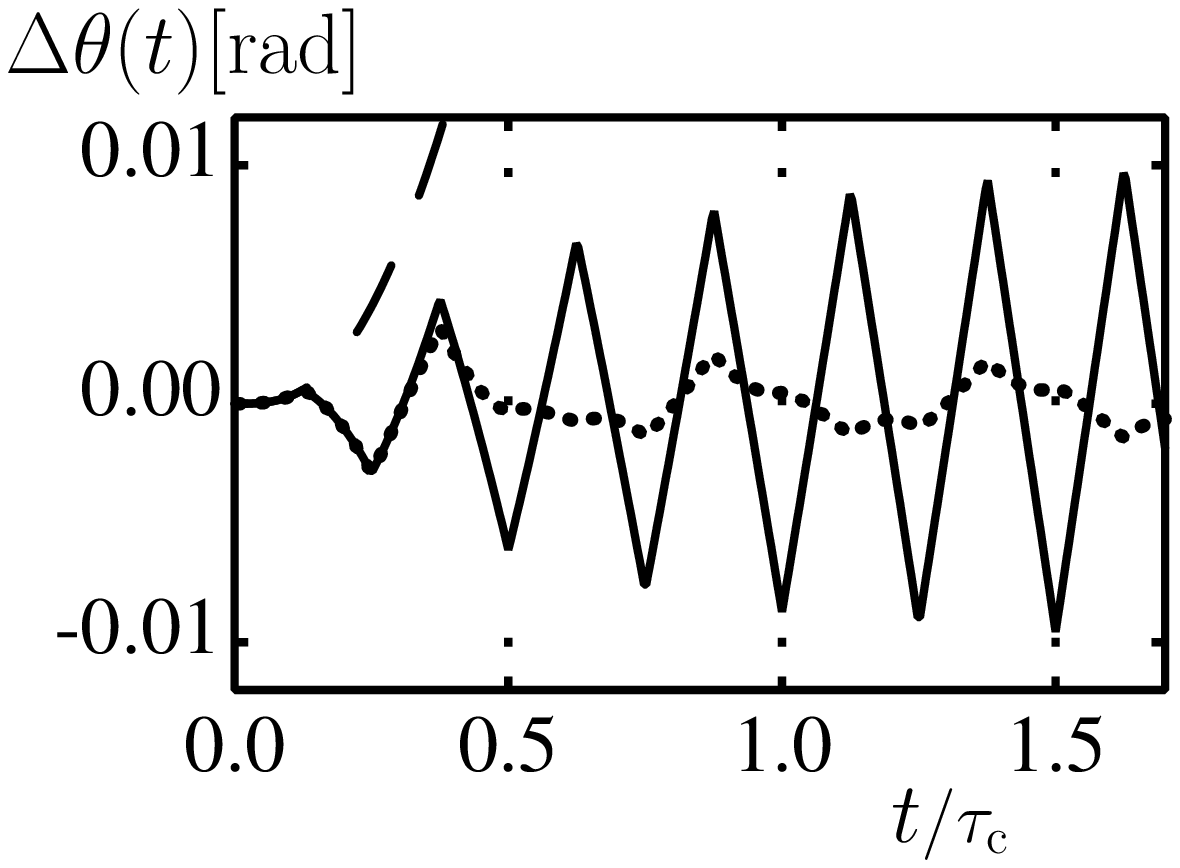}
\hspace{5mm}
{\Large(b)\hspace{-10mm}}
 \includegraphics*[scale=.63]{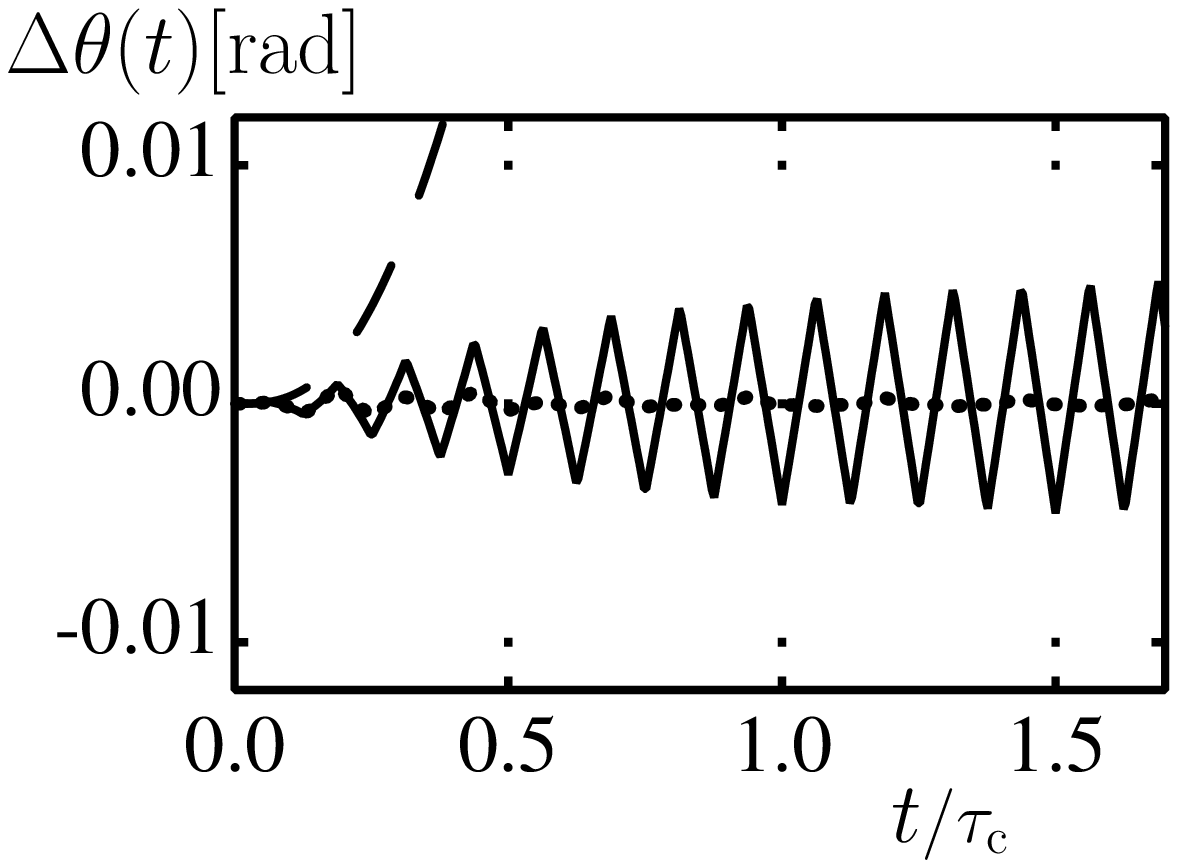}
\caption
 {
  Phase shift $\PShift(t)$ 
  in the case of $\tTimeNonRatio=2$
  with (a) $\Dt/\tTimeNon=2^{-3}$ and (b) $2^{-4}$.
  The solid lines and dotted lines correspond
  to the cases of the bb sequence and the bp sequence, respectively.
  The dashed lines represent the case where $\pi$ pulses are absent.
 }
\label{Phase}
\end{figure}

In Fig. \ref{Phase}, we display the phase shift $\PShift(t)$
for $\Dt/\tTimeNon=2^{-3}$ and $2^{-4}$
in the case of $\tTimeNonRatio=2$.
Obviously, $\PShift(t)\equiv0$
in the absence of qubit-environment interactions.
We observe that
the phase shift is suppressed by either of the bb or bp sequence.
The phase shift in the cases of $\tTimeNonRatio=5$ and $50$
is very small, so we do not display the corresponding results.

\section{Summary and Conclusion}
To study the effectiveness of the pulse control,
we have studied the time evolution of a qubit
coupled with a quantum environment under a periodic $\pi$ pulses.
We have considered a general interaction Hamiltonian
which induces not only the pure dephasing
but also the decoherence with population decay.
We have examined the effectiveness
of the bb sequence consisting of only bit-flips
and the bp sequence consisting of bit- and phase-flips.
We have derived the equation of motion for the reduced density matrix.
The resulting equation is applicable
to arbitrary sequences consisting of bit- and/or phase-flips.

By numerically solving the equation of motion,
we have shown that the effectiveness of the pulse sequences
depends on a relative strength of the two decoherence processes
especially when a pulse interval is slightly shorter than $\tTimeNon$.
We have found that
the bp sequence is more effective than, or at least as effective as,
the bb sequence regardless of $\tTimeNonRatio$
only in the short-interval limit.
This means that
if we cannot prepare periodic $\pi$ pulse with a short enough interval,
the bp sequence is not necessarily the best choice
unless $\tTimeNonRatio \sim 1$.

\section*{Acknowledgement}
The present authors thank Prof. M. Yamanishi
for calling their attention to ref. \citenum{ViolaLloyd}.

\end{document}